\documentstyle[preprint,aps]{revtex} 
\begin{document} 

\draft 

\preprint{UTPT-96-16} 

\title{Stochastic Gravity}
\author{J. W. Moffat} 

\address{Department of Physics, University of Toronto,
Toronto, Ontario, Canada M5S 1A7} 

\date{\today}

\maketitle 

\begin{abstract}%
Gravity is treated as a stochastic phenomenon based on fluctuations of the metric tensor
of general relativity. By using a (3+1) slicing of spacetime, a Langevin
equation for the dynamical conjugate momentum and a
Fokker-Planck equation for its probability distribution are derived.
The Raychaudhuri equation for a congruence of timelike or null geodesics leads to a stochastic
differential equation for the expansion parameter $\theta$ in terms of the proper time $s$. For
sufficiently strong metric fluctuations, it is shown that caustic singularities in spacetime can be
avoided for converging geodesics. The formalism is applied to the gravitational collapse
of a star and the Friedmann-Robertson-Walker cosmological model. It is found that
owing to the stochastic behavior of the geometry, the singularity in gravitational collapse
and the big-bang have a zero probability of occurring. Moreover, as a star collapses the
probability of a distant observer seeing an infinite red shift at the Schwarzschild
radius of the star is zero. Therefore, there is a vanishing probability of a Schwarzschild
black hole event horizon forming during gravitational collapse.

\end{abstract} 

\pacs{ } 

\section{Introduction}

Attempts to formulate a quantum theory of gravitation have not met with success
to date\cite{Isham}. The main attempts have been based on:

\begin{enumerate} 
\item Perturbation theory based on expansions about Minkowski spacetime;
\item Canonical Hamiltonian formalism;
\item Path integral formalism;
\item String theory.
\end{enumerate}

The first approach became untenable for two reasons: (a) The series expansions
were not renormalizable for gravitational couplings to matter and beyond the first
order for vacuum gravity. (b) The method chose a special frame of  reference, thus
failing to meet the fundamental requirements of general relativity. Approach (2)
ran into problems with the Hamiltonian constraints, since one set of constraints
was non-polynomial and could not be solved to yield the true gravitational degrees of
freedom. Attempts to resolve this problem have met with some success, but
fundamental difficulties still exist that prevent a satisfactory formulation of quantum
gravity\cite{Ashtekar}. In particular, it is difficult to meet the necessary
criterion that the quantum gravity theory must have a well-defined classical 
general relativity (GR) 
limit\cite{Smolin}. Approach (3) has met with the
unresolved difficulties of defining a meaningful measure in the spacetime manifold,
and the need to use an Euclidean spacetime to define the path integral. In general,
it is not possible to return to curved spacetime with a local Lorentzian signature by
a Wick rotation, and
guarantee that the classical GR limit exists. Finally, string theory\cite{Green}
has not succeeded in describing an easily interpretable solution to quantum gravity,
for the theory must be formulated as a non-perturbative scheme and potential problems
with causality and non-locality prevent the use of standard canonical Hamiltonian
methods and other field theory techniques based on perturbation theory.

Sakharov\cite{Sakharov} and Jacobson\cite{Jacobson} have formulated effective
gravitational theories. Sakharov viewed spacetime as an elastic bulk system with an
elasticity constant. However, this theory had difficulty yielding Newton's
and Einstein's gravitational theories in a well-defined limit. Jacobson pictured
Einstein's field equation as an equation of state and the gravitational variables
were thermodynamic variables, so that the Einstein equation was derived from the
proportionality of entropy and the black hole horizon area together with the
relation $\delta Q=TdS$ connecting heat $Q$, entropy $S$, and temperature $T$.
In this approach, it is not considered appropriate to quantize the classical gravitational
fields as is done for the case of the electromagnetic fields.

It is well known that an analogy between classical statistical mechanics and
Euclidean quantum field theory exists. This is based on the fact that the path integral
measure $D\phi\exp(-S_E[\phi])$ can be identified with the Boltzmann probability
distribution of a statistical system, where $S_E[\phi]$ denotes the field theory action,
which has been Wick rotated to a 4-dimensional Euclidean space. The Euclidean
Green functions are the correlation functions of a statistical system in thermal
equilibrium. A stochastic quantization of field theory has been formulated by
Parisi and Wu\cite{Parisi,Damgaard}. A new `fictitious' time coordinate was
introduced with respect to which all fields evolve, namely, the field $\phi(x)$ was
supplemented with the extra coordinate $t$, such that $\phi(x)\rightarrow \phi(x,t)$,
where $x^\mu=(x^0,x^1,x^2,x^3)$ and the fictitious time $t$ should not be confused
with the Euclidean time $x^0$. The evolution of $\phi(x,t)$ is determined by the
Langevin equation:
\begin{equation}
\partial_t\phi(x,t)=-\biggl[\frac{\delta S[\phi]}
{\delta\phi(x)}\biggr]_{\phi(x)=\phi(x,t)}+\xi_t(x,t),
\end{equation}
where $\xi_t(x,t)$ is a Markov process (Gaussian white noise). The basic assumption
of stochastic quantization is that in the limit $t\rightarrow \infty$ equilibrium is
reached, and that the equal time correlation functions $\phi[\eta]$ become identical
to the corresponding quantum Green functions:
\begin{equation}
{\rm lim}_{t\rightarrow \infty}\langle\phi[\eta](x_1,t_1)...\phi[\eta](x_n,t_n)\rangle
=\langle\phi(x_1)...\phi(x_n)\rangle,
\end{equation}
where the right hand-side denotes the standard vacuum expectation value in Euclidean
space. 

The stochastic quantization method has been used as a way to quantize
the gravitational field by Rumpf\cite{Rumpf}. However, since to any ordinary Feynman
diagram there corresponds a number of stochastic diagrams with the same topology,
and it can be shown that in the equilibrium limit these stochastic diagrams exactly sum up
to the corresponding Feynman diagram, then the standard problems of perturbative
quantum gravity will still prevail. Moreover, there is no indication that the method
alleviates any of the problems of non-perturbative quantum gravity.

Over the past few years, theoretical and experimental studies have shown
that a nonlinear system in a ``noisy" environment displays 
surprising behavior that does not conform to common intuitive experiences. 
Self-organization in non-equilibrium systems coupled to fluctuating environments
exhibits a much richer behavior than is possible in a deterministic
situation. Transition phenomena show features similar to equilibrium phase
transitions and to transition phenomena encountered in non-equilibrium systems 
influenced by external constraints, e.g., the B\'enard instability and the laser transition.
Phase transition phenomena and chaotic behavior can be induced by
environmental randomness\cite{Horsthemke}. Statistical mechanics can display
two kinds of systems: (1) a fine-tuning of parameters that leads to critical points
and phase transitions and (2), self-organization which can often describe critical
systems with structure spread out over every available scale\cite{Bak}. 
These properties can be applied to an understanding of our universe at both the large
and small scale\cite{Smolin2}. 

In quantum gravity there is only one scale that contains Planck's constant $\hbar$,
namely, the Planck length, $l_p=\sqrt{\hbar G}$, so that quantum gravity describes a
strongly coupled phase in which there are no correlations on large scales. However,
space and time can themselves be consequences of critical phenomena in which there occur
fluctuations with correlations on a scale
much larger than the Planck scale. Thus, there can exist a gravitational system
which is weakly coupled  and critical showing self-organization behavior. Such
gravitational phenomena can help to remove fine-tuning behavior encountered in
cosmology e.g., those associated with the initial conditions of the universe and
the cosmological constant. 

It is plausible that a nonlinear gravitational system coupled to
a stochastic, fluctuating spacetime geometry can lead to significant changes in the macroscopic
behavior of the system, for the fluctuations could be of order $V^0$ and not $V^{-1}$ where
$V$ denotes the characteristic volume of the system. Another way to ask this question is: do
nonlinear gravitational systems, coupled to a rapidly fluctuating spacetime geometry, always
adjust their macroscopic behavior to the average properties of the spacetime geometry or do there
exist situations in which the system responds to the randomness of the spacetime geometry,
displaying behavior forbidden under deterministic conditions? 

Attempts have been made to treat the geometry of spacetime in a semi-classical way by
writing Einstein's gravitational field equations in the form:
\begin{equation}
\label{Gequation}
G_{\mu\nu}=8\pi G\langle{\bar T}_{\mu\nu}\rangle,
\end{equation}
where $\langle {\bar T}_{\mu\nu}\rangle$ denotes the expectation value of the
stress-energy
tensor operator ${\bar T}_{\mu\nu}$ in a given quantum state. In order to reconcile
the left hand-side of (\ref{Gequation}) with the right hand-side, it is necessary to make
the gravitational field change in a discontinuous, acausal manner, a behavior unacceptable
from the classical point of view.

We shall develop a probabilistic interpretation of spacetime by defining the metric tensor
$g_{\mu\nu}$ as a stochastic random variable, which is defined by a set of possible
values and a probability distribution $P(g)$ over this set, with 
$P(g) \geq 0$ and normalized to unity on its range, where $g$ denotes the metric tensor
$g_{\mu\nu}$ and $P(g)$ is a scalar obtained from the contraction of a tensor quantity 
invariant under diffeomorphism transformations.
The probability that $g$ takes a value between $g$ and $dg$ is given by 
$P(g)dg$. We shall use the (3+1) foliation of spacetime in
GR\cite{Arnowitt,Misner,Wald,York} and the canonical formalism to derive a Langevin
equation for the conjugate canonical momentum, which is treated as a random variable. A
Fokker-Planck equation for the time evolution of this stochastic variable and an
equation for the probability density are derived. In this formalism, we treat both the
gravitational field and the matter in a probabilistic fashion, so that Eq.(\ref{Gequation})
now becomes
\begin{equation}
\label{Geq}
\langle G_{\mu\nu}\rangle=8\pi G\langle T_{\mu\nu}\rangle,
\end{equation}
and there is no logical conflict between the two sides of the equation, although the
$G_{\mu\nu}$ and the $T_{\mu\nu}$ are not treated as quantum mechanical
operators but instead are considered to be functions of random, stochastic variables.

In Sections II through V, we introduce basic
notation for stochastic theory and probability as applied to GR, and in Section VI,
we derive the Langevin equation for the conjugate momentum variable in $(3+1)$
foliated spacetime and a Fokker-Planck equation for the probability density. In Section
VII, we use the Raychaudhuri equation to derive a stochastic differential equation
for the expansion parameter $\theta$, and investigate the stochastic behavior of
a converging congruence of geodesics. In sections VIII and IX, we apply the
method to gravitational collapse and cosmology and we conclude, in Section X, with
a summary of the results.

\section{Metric Fluctuations and Stochastic Processes}

Wheeler\cite{Misner} suggested that quantum fluctuations in the metric
tensor of spacetime, $g_{\mu\nu}$, should occur at the order of the Planck length:
$l_P=({\hbar}G)^{1/2}=1.6\times 10^{-33}{\rm cm}$. These fluctuations are of
order:
\begin{equation}
\Delta g\sim\frac{l_P}{L},
\end{equation}
where $L$ is the spatial dimension of the region in a local Lorentz frame of reference.
Fluctuations in the curvature tensor in space are of the order:
\begin{equation}
\Delta R\sim \frac{\Delta g}{L^2}\sim \frac{l_P}{L^3}.
\end{equation}
For a domain of order $1\, {\rm cm}$, the curvature fluctuations are
\begin{equation}
\Delta R\sim 10^{-33}{\rm cm}^{-2},
\end{equation}
and for subatomic domains the metric fluctuations are $\Delta g\sim 10^{-20}$. These
are so small that we can ignore them for experimentally accessible physical observations.

Let us postulate that the spacetime geometry is a physical system. In classical GR, it is assumed
that the spacetime manifold is $C^2$ smooth down to zero length scales. Since all known
physical systems possess ``noise" at some length scale, the former assumption would seem to be
unrealistic. We shall also postulate that the spacetime geometry has an (unknown) underlying
microscopic structure and that the subsystems of this structure undergo stochastic fluctuations
at a certain length scale, which can produce macroscopic correlation lengths due to cooperative,
self-organized behavior when the subsystems couple to matter.

If we treat the geometry of spacetime as a fluctuating environment, then
we can consider the gravitational constant $G$ as a control parameter with the
decomposition:
\begin{equation}
\label{control parameter}
G_t=G+\sigma\xi_t,
\end{equation}
where $G$ is the Newtonian gravitational constant ( $G=6.673\times 10^{-8}
{\rm g}^{-1}\,{\rm cm}^3\,{\rm s}^{-2}$ ), and $\xi_t$ is
Gaussian white noise with $E\{\xi_t\}=0$ where $E$ denotes the expectation
value; $\sigma$ measures the intensity of the
geometrical fluctuations of the metric. Thus, $G$ will
have a bell-shaped curve distribution, peaked at the average value of
$G$. The central limit theorem assures that the fluctuations in $G$ are Gaussian
distributed. The spacetime fluctuations of the metric are of order:
\begin{equation}
\Delta g\sim \frac{\sigma\xi_tM}{L}.
\end{equation}

Let us describe the gravitational system by a set of state variables $X$ obeying
the equation\cite{Gihman,Oksendal,Horsthemke}:
\begin{equation}
\partial_tX(r,t)=f_G(X(r,t)),
\end{equation}
which describes the local evolution of $X$ in space $r$ and time $t$. The state variables $X(t)$
are constructed from the metric tensor $g$, its derivatives and possible curvature quantities, and
the functionals $f_G(X)$ depend on the control parameter $G$.
When the symmetry of the gravitating system is broken, self-organization sets in and
the disorganized state becomes unstable.  We assume that there exists a time
independent solution ${\bar X}$ satisfying
\begin{equation}
f_G({\bar X})=0,
\end{equation}
corresponding to an unorganized state. We expand $X$ about ${\bar X}$ in small
perturbations:
\begin{equation}
X(r,t)={\bar X}+x(r,t),
\end{equation}
where $\vert x/{\bar X}\vert \ll 1$, and the perturbations $x(r,t)$ satisfy the equations:
\begin{equation}
\partial_tx_i=\sum_kB_{ik}x_k,
\end{equation}
with the solutions:
\begin{equation}
x_i(r,t)=\sum_kx_i^k(r)\exp(\omega_kt).
\end{equation}

The life-time of gravitational disturbances is of order
\begin{equation}
\tau_{\rm macro}=\left\vert\frac{1}{\rm Re(\omega_k)}\right\vert,
\end{equation}
where $\tau_{\rm macro}$ is the macroscopic time scale of evolution of the 
system.  The ${\rm Re}(\omega_k)$ must be negative for a gravitational state ${\bar X}$
which is asymptotically stable. Gravitational transitions can occur at the point
$G=G_c$ at which ${\rm Re}(\omega_k)$ changes sign from negative to positive values,
and a new solution can occur at this point of bifurcation associated with a gravitational
order parameter which vanishes at $G_c$.

The correlation time $\tau_{\rm cor}$ is a measure of the rapidity of the random
fluctuations of the spacetime geometry. For a stationary gravitational process it is
defined by
\begin{equation}
\tau_{\rm cor}=\frac{1}{C(0)}\int_0^{\infty}C(\tau)d\tau,
\end{equation}
where the normalized correlation function is given by
\begin{equation}
{\tilde C}(\tau)=\frac{E\{\delta g_t\delta g_{t+\tau}\}}{E\{\delta g_t^2\}}.
\end{equation}
For a typical exponentially decreasing correlation function,
\begin{equation}
C(\tau)=C(0)\exp(-\gamma\tau),
\end{equation}
where $\gamma$ is a constant, $\tau_{\rm cor}=1/\gamma$.
Spacetime will be characterized by rapid fluctuations with
\begin{equation}
\tau_{\rm cor} \ll \tau_{\rm macro}.
\end{equation}

For a given 3-geometry ${}^{(3)}{\cal G}$, including geometrical noise in the
gravitational equations leads to the formal
Langevin\cite{Langevin} or stochastic differential equation (SDE):
\begin{equation}
\label{dotgeq}
\partial_tX(g_t)=f_{G_t}(g_t),
\end{equation}
where $g_t$ denotes the random metric tensor. 
Assuming for the moment that the external parameter $G_t$ enters in a linear way,
then (\ref{dotgeq}) takes the form:
\begin{equation}
\label{SDE}
\partial_tX=f(g_t)+Gh(g_t)+\sigma\xi_th(g_t).
\end{equation}
The extreme irregularity of Gaussian white noise means that the time derivative of the
metric is not well defined. However, the standard theory of stochastic processes can
handle this difficulty by defining the equivalent integral equation:
\begin{equation}
X_t=X_0+\int^t_0F_{G_t}(g_y)dy+\sigma\int^t_0h(g_y)\xi_ydy,
\end{equation}
where
\begin{equation}
F_{G_t}(g_t)=f(g_t)+Gh(g_t).
\end{equation}
Two definitions of the stochastic integral $\int h(g_y)\xi_ydy$ have been given by
Ito\cite{Ito} and Stratonovich\cite{Stratonovich}. Both of these definitions are
based on the concept that integration of Gaussian white noise yields Brownian
motion, which we denote by $W_t$, i.e., $\xi_t=\partial_tW_t$. Then,
\begin{equation}
\int h(g_y)\xi_ydy=\int h(g_y)dW_y.
\end{equation}
Both the Ito and the Stratonovich definitions of the integral lead to a consistent
calculus.

\section{Elements of Metric Fluctuation Probability Theory}

For a manifold ${\cal M}$ in 4-space, we assume the existence of a probability
triple $(\Omega, {\cal A},P)$ consisting of a sample space $\Omega$, a field of events
${\cal A}$ and a probability measure $P$\cite{Pitman}. Only $\Omega$ 
and ${\cal A}$ are used
to define a random variable in ${\cal M}$. The sample space $\Omega$ is the ensemble
of elementary outcomes of spacetime measurements in ${\cal M}$, labeled
$\omega:\omega\in \Omega$. We take it as given that the number of elementary outcomes
in the manifold ${\cal M}$ is uncountably infinite.

The second element of the probability triple of ${\cal M}$ is ${\cal A}$ (the $\sigma$
field) of events. The set ${\cal A}$ satisfies the following properties:
{\obeylines\smallskip
(i) ${\cal A}$ contains the certain event, $\Omega\in {\cal A}$.

(ii) It contains the empty set $\phi$: $\phi\in {\cal A}$.

(iii) If $A$ is an event, then so also is the complement $\bar A=\Omega-A$.

(iv) If $A$ and $B$ are events, so is their union and intersection.
\smallskip}

The metric tensor $g$ is defined as a random variable by the requirement that it is a
function from the sample space $\Omega$ into some state space: $g:\Omega\rightarrow
{\cal R}$. We also have
\begin{equation}
A=X^{-1}((-\infty,g])\in {\cal A},\quad \forall g\in {\cal R},
\end{equation}
where 
\begin{equation}
X^{-1}(B)=\{\omega\vert g(\omega)\in B\},\quad B \subset {\cal R}.
\end{equation}

We define the third element $P$ as the frequency of occurrence of event $A$.
Thus, on the interval $[0,1]$ $P: {\cal A}\rightarrow [0,1]$. We have by definition:
\begin{mathletters}
\begin{eqnarray}
P(\phi)&=&0,\\
P(\Omega)&=&1,\\
A_n\in {\cal A},\quad A_n\cap_{n\not=m}A_m
&=&\phi\Rightarrow P\biggl(\cup_{n=1}^{\infty}
A_n\biggr)=\sum^{\infty}_{n=1}P(A_n).
\end{eqnarray}
\end{mathletters}

The distribution function induced by the random metric variable $g$ is defined by
\begin{equation}
F_g=P(\{\omega\vert g(\omega)\})=P(g),\quad g\in {\cal R}.
\end{equation}
The probability density $p(g)$ is defined by
\begin{equation}
F(g)=\int^gp(y)dy,
\end{equation}
where the integral is defined as a functional integral over the metric $g$ with a given
measure. In the following, it is understood that the distribution function $F_g$ and the
probability density $p(g)$ transform as tensor quantities under diffeomorphism
transformations of the coordinates.

In the standard way, we define the expectation $E\{g\}$ of the metric tensor by
\begin{equation}
E\{g\}=\int_{\Omega}gdP=\int_{\Omega}g(\omega)dP(\omega)
=\int_{\Omega}g(\omega)P(d\omega),
\end{equation}
and we have
\begin{equation}
E\{g\}=\int_{{\cal R}}gp(g)dg,
\end{equation}
where $E\{g\}$ is a tensor quantity. The rth moment of $g$ is
\begin{equation}
E\{g^r\}=\int_{\Omega}g^r(\omega)dP(\omega)=\int_{{\cal R}}g^rdF(g)
=\int_{{\cal R}}g^rp(g)dg.
\end{equation}
The mean deviation of the random metric function is
\begin{equation}
\sigma^2=E\{(\delta g)^2\}=E\{(g-m)^2\}=\int_{{\cal R}}(g-m)^2dF(g),
\end{equation}
where $m$ denotes the mean value of $g$. The number and location of the extrema of the
probability density are
\begin{equation}
\partial_gp(g)\vert_{g=g_m}=0,
\end{equation}
where $\partial_gp(g)$ denotes a functional derivative with respect to the metric $g$.
The most probable states are given by the maxima:
\begin{equation}
\partial_{gg}p(g)\vert_{g=g_m} < 0.
\end{equation}

We shall denote stochastic processes for the metric tensor in a 3-geometry
${}^{(3)}{\cal G}$ by $g_t$ and deterministic
metric processes by $g$. The set $\{g_t;t\in\theta\}$ of real-valued random metric
variables, $g_t:(\Omega,{\cal A},P)\rightarrow ({\cal R},{\cal B})$, describes a
random function with the index $\theta$ and a state space ${\cal R}$. The index
parameter will be the time and the index set $\theta$ is either the real line ${\cal R}$
or the non-negative half-line, if the process started at $t=0$.

The standard Wiener process $W_t$ has the intial condition: $W_0=0$. The
hierarchy of probability densities for a one-dimensional fluctuating metric is
\begin{mathletters}
\begin{equation}
p(g,t)=(2\pi t)^{-1}\exp(-g^2/2t)\equiv n(g,t),
\end{equation}
\[
\vdots
\]
\begin{equation}
p(g_1,t_1;...;g_m,t_m)=n(g_1,t_1)n(g_2-g_1,t_2-t_1)...n(g_m-g_{m-1},t_m-t_{m-1}).
\end{equation}
\end{mathletters}

Since the Wiener process is a Gaussian process, we define the finite dimensional
probability density as Gaussian:
\begin{equation}
p_G(g)=[(2\pi)^n{\rm det}M]^{-1/2}\exp[-1/2(g-m)^{tr}M^{-1}(g-m)],
\end{equation}
where
\begin{equation}
g^{tr}=(g_1,...,g_n),\quad m^{tr}=(m_1,...,m_n),
\end{equation}
and $M$ is a positive definite $n\times n$ matrix.

A {\it stationary} stochastic process in a 3-geometry ${}^{(3)}{\cal G}$ has a
probability density that is invariant against shifts of time:
\begin{equation}
p(g_1,t_1;...;g_n,t_n)=p(g_1,t_1+t;...;g_n,t_n+t).
\end{equation}
Then, we have
\begin{equation}
p(g,t)=p_S(g),
\end{equation}
and
\begin{equation}
E\{g_t\}=\int_{{\cal R}}gp(g,t)dg=\int_{{\cal R}}gp_S(g)dg=m.
\end{equation}
Moreover, the two-dimensional probability density satisfies
\begin{equation}
p(g_1,t_1;g_2,t_2)=p(g_1,g_2;t_2-t_1),
\end{equation}
so it only depends on the time difference $t_2-t_1$. The correlation function,
defined by
\begin{equation}
E\{\delta g_{t_1}\delta g_{t_2}\}=\int_{{\cal R}}\int_{{\cal R}}(g_1-m)
(g_2-m)p(g_1,t_1;g_2,t_2)dg_1 dg_2,
\end{equation}
also only depends on the time difference $t_2-t_1$.

The Gaussian white noise is very irregular but is a useful model for
rapidly fluctuating phenomena. Since it has no continuous sample paths and it has infinite
total power:
\begin{equation}
S=2\int^{\infty}_0({\bar\sigma}^2/2\pi)d\nu=\infty,
\end{equation}
it does not occur in nature. In the idealization of $\delta$-function correlated 
metric fluctuations, the gravitational system will be described by an SDE or Langevin
equation of the form (\ref{SDE}), where $\sigma$ denotes the intensity of the Gaussian
white noise.

A random process for the metric fluctuations is a Markov process if when the present
state of the process $g_t$ is known, then any additional information on its past history
is irrelevant for the prediction of its future evolution. The only Markov processes
which have continuous and differentiable sample paths are the deterministic ones
associated with classical gravity theory, given in our ${}^{(3)}{\cal G}$ by
\begin{equation}
\partial_tX(g(t))=f(g(t)),
\end{equation}
with $X_{t_0}=I$, where the only random element is the initial value $I$. For 
every positive $\epsilon$ and every function of the metric, $f(g,\tau)$, there is a drift,
\begin{equation}
{\rm lim}_{t\rightarrow \tau}\int_{\vert y-g\vert\leq\epsilon}
p(y,t\vert g,\tau)dy=f(g,\tau).
\end{equation}
For every positive $\epsilon$, there is a function $h(g,\tau)$ called diffusion, which
satisfies
\begin{equation}
{\rm lim}_{t\rightarrow\tau}\int_{\vert y-g\vert\leq\epsilon}(y-g)^2p(y,t\vert g,\tau)dy=
h^2(g,\tau).
\end{equation}
We shall also assume that there exists a positive $\delta$ such that the condition:
\begin{equation}
{\rm lim}_{t\rightarrow\tau}\int_{\cal R}\vert y-g\vert^{2+\delta}p(y,t\vert g,\tau)=0,
\end{equation}
is satisfied.

\section{The Fokker-Planck Equation and Stationary Probabilities}

The transition probability density $p(y,t\vert g,\tau)$ of the metric diffusion
process $g_t$ satisfies the Fokker-Planck equation\cite{Fokker,Planck}(FPE) or
Kolmogorov forward equation\cite{Gihman}:
\begin{equation}
\label{Fokkereq}
\partial_tp(y,t\vert g,\tau)=-\partial_y[f(y,t)p(y,t\vert g,\tau)]
+\frac{1}{2}\partial_{yy}[h^2(y,t)p(y,t\vert g,\tau)].
\end{equation}
Here, it is assumed that the functional derivatives occurring in (\ref{Fokkereq}) exist and
are continuous. As in the previous sections, we have suppressed tensor indices.
For time homogeneous Markov metric processes, the Fokker-Planck equation is given by
\begin{equation}
\partial_tp(y,t\vert g)=-\partial_y[f(y)p(y,t\vert g)]+\frac{1}{2}\partial_{yy}
[h^2(y)p(y,t\vert g)].
\end{equation}
Here, the drift and the diffusion are time independent. For a one-dimensional system, the
FPE for the probability density becomes
\begin{equation}
\partial_tp(g,t)=-[\partial_gf(g,t)p(g,t)]+\frac{1}{2}\partial_{gg}[h^2(g,t)p(g,t)].
\end{equation}

An important and useful solution of the FPEs can be obtained for stationary random
metric processes. We expect that a gravitational system subjected to metric fluctuations
for a sufficiently long time will settle down to a stationary behavior. This means that as
time approaches infinity the system will reach a state for which the probability density
$p_S(g)$ has a shape that does not change with time, i.e., the probability to find the
system in the neighborhood of a particular state becomes time independent. However,
the sample paths $g_t(\omega)$ will in general not approach a steady-state value
$g_S(\omega)$, so that the state of the system still fluctuates. However, these
fluctuations are such that $g_t$ and $g_{t+\tau}$ have the same probability density,
$p_S(g)$.

The stationary solution $p_S(g)$ of the FPE satisfies:
\begin{equation}
\partial_tp(g,t\vert g_0,0)+\partial_gI(g,t\vert g_0,0)=0,
\end{equation}
where
\begin{equation}
I(g,t\vert g_0,0)=f(g)p(g,t\vert g_0,0)-\frac{\sigma^2}{2}
\partial_g[h^2(g)p(g,t\vert g_0,0)],
\end{equation}
and we have for simplicity considered the time homogeneous case. The stationary FPE
is then given by
\begin{equation}
\partial_gI_S(g)=0,
\end{equation}
which implies that $I_S(g)={\rm constant}$ for $g\in[g_1,g_2]$.

The solution of the stationary FPE equation reads\cite{Horsthemke}
\begin{equation}
p_S(g)=\frac{C}{h^2(g)}\exp\biggl(\frac{2}{\sigma^2}
\int^g\frac{f(x)}{h^2(x)}dx\biggr)
-\frac{2}{\sigma^2h^2(g)}
I\int^g\exp\biggl(\frac{2}{\sigma^2}\int^g_y\frac{f(x)}{h^2(x)}dx\biggr)dy,
\end{equation}
where $C$ denotes a normalization constant.
When the boundaries of the gravitational system are natural, i.e., regular with
instantaneous reflection imposed as a boundary condition, then there is no
flow of probability out of the state space and $I=0$.  By using the Ito
prescription, we obtain
\begin{equation}
p_S(g)=\frac{C}{h^2(g)}\exp\biggl(\frac{2}{\sigma^2}\int^g
\frac{f(x)}{h^2(x)}dx\biggr).
\end{equation}
In this case, the normalization constant $C$ is given by
\begin{equation}
C^{-1}=\int_{g_1}^{g_2}\frac{1}{h^2(g)}\exp\biggl(\frac{2}{\sigma^2}
\int^g\frac{f(x)}{h^2(x)}dx\biggr)dg <\infty.
\end{equation}

\section{Nonlinear Gravitational Stochastic Systems}

If the gravitational equations are linear in the control parameter $G$, then we can
apply directly the stochastic methods to solve for the probability density functions.
However, we shall find that in certain applications the gravitational constant will enter
the phenomenological equations in a nonlinear way. Consider a phenomenological
equation of the form:
\begin{equation}
\partial_tX(t)=f(X(t))+\beta(G)h(X(t)),
\end{equation}
where $\beta$ is a nonlinear function of the gravitational constant $G$. For the
geometrical fluctuations, we replace the constant control parameter $G$ by a
stationary stochastic process $G_t=G+\zeta_t$ and form the SDE:
\begin{equation}
\label{nonlinearSDE}
dX_t=[f(X_t)+\beta(G+\zeta_t)h(X_t)]dt.
\end{equation}
Since nonlinear operations on generalized functions, such as the Dirac $\delta$-function,
cannot be given a well-defined mathematical meaning, we cannot use a white-noise
Gaussian approximation by setting $G_t=G+\sigma\xi_t$. However, methods exist
which lead to satisfactory approximations to the white-noise 
idealization\cite{Miguel,Horsthemke}. 

We shall assume that $\zeta_t$ is ``colored" noise and that in applications the geometry
of spacetime varies on a much faster time scale than the gravitating matter system 
coupled to it, so that $\zeta_t$ is a process with a short correlation time. Let us define the
geometrical fluctuations by the process $\eta_t$:
\begin{equation}
\eta_t=\beta(G+\zeta_t)-E\{\beta(G+\zeta_t)\}=\beta(G+\zeta_t)-m(G,\sigma^2).
\end{equation}
Then we can write Eq.(\ref{nonlinearSDE}) as
\begin{equation}
dX_t=[f(X_t)+m(G,\sigma^2)h(X_t)]dt+h(X_t)\eta_tdt.
\end{equation}
We can ``speed up" the stochastic process by writing
\begin{equation}
\eta_t^\epsilon=\eta_{t/\epsilon^2}=\beta(G+\zeta_{t/\epsilon^2})-m(G,\sigma^2),
\end{equation}
where $\epsilon=\tau_{\rm cor}$ is a small parameter. Then we have
\begin{equation}
dX_t^\epsilon=[f(X_t^\epsilon)+m(G,\sigma^2)h(X_t^\epsilon)]dt
+\frac{1}{\epsilon}h(X_t)\eta_t^\epsilon dt,
\end{equation}
and
\begin{equation}
d\zeta_t=-\frac{1}{\epsilon^2}\zeta_tdt+\frac{\sigma}{\epsilon}dW_t.
\end{equation}

Rescaling time, $t\rightarrow t/\tau_{\rm cor}$, the FPE has the form:
\begin{equation}
\partial_tp(x,y,t)=\biggl(\frac{D_1}{\epsilon}+\frac{D_2}{\epsilon}+D_3\biggr)
p^\epsilon(x,y,t),
\end{equation}
where
\begin{mathletters}
\begin{eqnarray}
D_1&=&\partial_yy+\frac{\sigma^2}{2}\partial_yy,\\
D_2&=&-y\partial_xh(x),\\
D_3&=&-\partial_x[f(x)+m(G,\sigma^2)h(x)].
\end{eqnarray}
\end{mathletters}
The correct SDE in the white-noise Gaussian limit is given by
\begin{equation}
dX_t=[f(X_t)+m(G,\sigma^2)h(X_t)]dt+{\tilde\sigma}h(X_t) dW_t,
\end{equation}
where 
\begin{equation}
{\tilde\sigma}^2=2\int_{\cal R}p_S(y)[\beta(G+y)-m(G,\sigma^2)]
{\tilde\beta}(G+y)dy,
\end{equation}
and $p_S(x)$ is the stationary probability density, which has the form:
\begin{equation}
p^\epsilon_S(x)=\frac{C}{v(x)^2}\exp\biggl(\frac{2}{\sigma^2}\int 
\frac{F(x)}{v(x)h(x)}dx\biggr),
\end{equation}
where
\begin{equation}
F(x)=f(x)+m(G,\sigma^2)h(x)
\end{equation}
and
\begin{equation}
v(x)=h(x)[1+\epsilon Q_1(x)+\epsilon^2Q_2(x)+...].
\end{equation}

\section{Hamiltonian Formulation of Gravity and Stochastic Differential Equations}

We shall adopt the standard Lagrangian density in Einstein's gravitational theory:
\begin{equation}
{\cal L}={\cal L}_G+{\cal L}_M,
\end{equation}
where
\begin{equation}
{\cal L}_G=\sqrt{-g}g^{\mu\nu}R_{\mu\nu},
\end{equation}
${\cal L}_M$ is the Lagrangian density for the matter field, $g={\rm det}\,g_{\mu\nu}$ and 
$R_{\mu\nu}$ is the Ricci curvature tensor.

The control parameter is defined in terms of spacetime fluctuations of the gravitational constant:
\begin{equation}
G_x=G+\sigma\xi_x,
\end{equation}
where, as in Eq. (\ref{control parameter}), $G$ is the average value of the gravitational constant,
and $\xi_x$ denotes Gaussian spacetime noise. The diffeomorphism invariant Einstein field
equations including stochastic spacetime fluctuations take the form:
\begin{equation}
G_{\mu\nu}=8\pi(GT_{\mu\nu}+\sigma\xi_xT_{\mu\nu}),
\end{equation}
where $G_{\mu\nu}$ is the Einstein tensor: $G_{\mu\nu}=
R_{\mu\nu}-\frac{1}{2}g_{\mu\nu}R$ and $R$ is the scalar curvature.

The first step in formulating the Hamiltonian approach to gravity is to introduce a
foliation of spacetime which defines a 3-geometry ${}^{(3)}{\cal G}$. We choose a time
function $t$ and a vector field $t^\mu$ such that the surfaces, $\Sigma$, of constant
$t$ are spacelike Cauchy surfaces with $t^\mu\nabla_\mu t=1$, where $\nabla_\mu$
denotes the covariant differentiation with respect to the metric tensor $g_{\mu\nu}$.
In contrast to Minkowski spacetime, there is no preferred coordinate system in
curved spacetime.  In the standard $(3+1)$ treatment of 
spacetime\cite{Arnowitt,Misner,Wald,York}, we introduce the lapse function, $N$, by
\begin{equation}
N=-g_{\mu\nu}t^\mu n^\nu=(n^\mu\nabla_\mu t)^{-1}
\end{equation}
and the shift vector $N^\mu$ by
\begin{equation}
N^\mu={h^\mu}_\nu t^\nu,
\end{equation}
where $n^\mu$ is the unit normal vector to $\Sigma$,
\begin{equation}
h_{\mu\nu}=g_{\mu\nu}+n_\mu n_\nu,
\end{equation}
is the induced spatial metric on $\Sigma$, and we use the metric signature: $(-1, +1, +1,
+1)$.
We have
\begin{equation}
n^\mu=\frac{1}{N}(t^\mu-N^\mu).
\end{equation}

In the following, the volume element is
\begin{equation}
\sqrt{-g}=N\sqrt{h}.
\end{equation}
The scalar curvature, $R$, is given by
\begin{equation}
R=2(G_{\mu\nu}n^\mu n^\nu-R_{\mu\nu}n^\mu n^\nu),
\end{equation}
We also have
\begin{equation}
R_{\mu\nu}n^\mu n^\nu={R_{\mu\sigma\nu}}^\sigma n^\mu n^\nu=
K^2-K_{\mu\nu}K^{\mu\nu}
-\nabla_\mu(n^\mu\nabla_\sigma n^\sigma)+\nabla_\sigma(n^\mu\nabla_\mu n^\sigma),
\end{equation}
where $K_{\mu\nu}$ is the extrinsic curvature of $\Sigma$ defined by
\begin{equation}
K_{\mu\nu}=\frac{1}{2}N^{-1}[\partial_th_{\mu\nu}-D_\mu N_\nu-D_\nu N_\mu],
\end{equation}
and $D_\mu$ is the derivative operator on the surface $\Sigma$ connected with
$h_{\mu\nu}$.  This leads to the expression for ${\cal L}_G$ obtained by Arnowitt,
Deser, and Misner\cite{Arnowitt}:
\begin{equation}
{\cal L}_G=\sqrt{h}N[{}^{(3)}R+K_{\mu\nu}K^{\mu\nu}-K^2].
\end{equation}

The Hamiltonian density associated with ${\cal L}_G$ is given by
\begin{eqnarray}
\label{Hamiltonian}
{\cal H}_G&=&\pi^{\mu\nu}\partial_th_{\mu\nu}-{\cal L}_G=
\sqrt{h}\biggl\{N\biggl[-{}^{(3)}R+h^{-1}\pi^{\mu\nu}\pi_{\mu\nu}
-\frac{1}{2}h^{-1}{\pi^\lambda}_\lambda{\pi^\sigma}_\sigma\biggr]\nonumber\\
&&\mbox{}
-2N_\nu[D_\mu(h^{-1/2}\pi^{\mu\nu})]
+2D_\mu(h^{-1/2}N_\nu\pi^{\mu\nu})\biggr\},
\end{eqnarray}
where $\pi^{\mu\nu}$ is the canonically conjugate momenta to $h_{\mu\nu}$ defined by
\begin{equation}
\pi^{\mu\nu}=\frac{\partial {\cal L}_G}{\partial (\partial_th_{\mu\nu})}
=\sqrt{h}(K^{\mu\nu}-h^{\mu\nu}K).
\end{equation}
In (\ref{Hamiltonian}), the $N$ and $N_\mu$ are not
treated as dynamical variables but, instead, define the configuration space in terms of the
metric, $h_{\mu\nu}$. 

Variation of 
\begin{equation}
H_G=\int_{\Sigma}{\cal H}_G,
\end{equation}
leads to the constraint equations:
\begin{mathletters}
\begin{eqnarray}
\label{1constraint}
{}^{(3)}R-h^{-1}\pi^{\mu\nu}\pi_{\mu\nu}+\frac{1}{2}h^{-1}
{\pi^\mu}_\mu{\pi^\sigma}_\sigma&=&16\pi GT_{\bot \bot},\\
\label{3constraints}
D_\mu(h^{-1/2}\pi^{\mu\nu})&=&-8\pi GT^\nu_{\bot},
\end{eqnarray}
\end{mathletters}
where $T_{\bot\bot}=T_{\mu\nu}n^\mu n^\nu$ and $T^\nu_{\bot}={h^\nu}_\alpha
T^{\alpha\beta}n_\beta$ are components of the stress energy-momentum tensor $T_{\mu\nu}$
for matter.

Hamilton's first order equations in time are now obtained from $H_G$\cite{Arnowitt}:
\begin{mathletters}
\begin{eqnarray}
\label{heq}
\partial_th_{\mu\nu}&=&\frac{\delta H_G}{\delta\pi^{\mu\nu}}
=2h^{-1/2}N\biggl(\pi_{\mu\nu}-\frac{1}{2}h_{\mu\nu}{\pi^\sigma}_\sigma\biggr)
+2D_{(\mu}N_{_\nu)},\\
\label{pieq}
\partial_t\pi^{\mu\nu}&=&-\frac{\delta H_G}{\delta h_{\mu\nu}}
=-Nh^{1/2}\biggl({}^{(3)}R^{\mu\nu}-\frac{1}{2}{}^{(3)}Rh^{\mu\nu}\biggr)
+\frac{1}{2}Nh^{-1/2}h^{\mu\nu}\biggl(\pi^{\sigma\tau}\pi_{\sigma\tau}
-\frac{1}{2}{\pi^\lambda}_\lambda{\pi^\sigma}_\sigma\biggr)\nonumber\\
& & \mbox{}
-2Nh^{-1/2}\biggl(\pi^{\mu\sigma}{\pi^\nu}_\sigma
-\frac{1}{2}{\pi^\sigma}_\sigma\pi^{\mu\nu}\biggr)
+h^{1/2}(D^\mu D^\nu N-h^{\mu\nu}D^\sigma D_\sigma N)\nonumber\\
&&\mbox{}
+h^{1/2}D_\sigma(h^{-1/2}N^\sigma\pi^{\mu\nu}
-2\pi^{\sigma(\mu}
D_\sigma N^{\nu)})+8\pi GT^{\mu\nu}.
\end{eqnarray}
\end{mathletters}

We shall now write the stochastic differential equation for the dynamical random variable
$\pi_t^{\mu\nu}$ as
\begin{equation}
\label{pisde}
\partial_t\pi_t^{\mu\nu}
=f_t^{\mu\nu}(\pi_t)+8\pi GT^{\mu\nu}+8\pi\sigma\xi_tT^{\mu\nu},
\end{equation}
where
\begin{eqnarray}
f_t^{\mu\nu}(\pi_t)&
=&-Nh^{1/2}\biggl({}^{(3)}R^{\mu\nu}-\frac{1}{2}{}^{(3)}Rh^{\mu\nu}\biggr)
+\frac{1}{2}Nh^{-1/2}h^{\mu\nu}\biggl(\pi^{\sigma\tau}\pi_{\sigma\tau}
-\frac{1}{2}{\pi^\mu}_\mu{\pi^\sigma}_\sigma\biggr)\nonumber\\
&&\mbox{}
-2Nh^{-1/2}\biggl(\pi^{\mu\sigma}{\pi^\nu}_\sigma
-\frac{1}{2}{\pi^\lambda}_\lambda\pi^{\mu\nu}\biggr)
+h^{1/2}(D^\mu D^\nu N-h^{\mu\nu}D^\sigma D_\sigma N)\nonumber\\
&&\mbox{}
+h^{1/2}D_\sigma(h^{-1/2}N^\sigma\pi^{\mu\nu}
-2\pi^{\sigma(\mu}D_\sigma N^{\nu)}).
\end{eqnarray}
Exploiting the fact that the stochastic equation is the derivative of the Wiener process,
and using the control parameter, $G_t=G+\sigma\xi_t$, we obtain
\begin{equation}
\label{piwiener}
d\pi_t^{\mu\nu}=F_t^{\mu\nu}(\pi_t)dt+B_t^{\mu\nu}dW_t,
\end{equation}
where
\begin{equation}
F_t^{\mu\nu}(\pi_t)=f_t^{\mu\nu}(\pi_t)+8\pi GT^{\mu\nu},
\end{equation}
and
\begin{equation}
B_t^{\mu\nu}=8\pi\sigma T^{\mu\nu}.
\end{equation}

The constraint equations (\ref{1constraint}) and (\ref{3constraints}) take the stochastic form:
\begin{mathletters}
\begin{eqnarray}
{}^{(3)}R_t-h_t^{-1}\pi_t^{\mu\nu}\pi_{t\,\mu\nu}+\frac{1}{2}h_t^{-1}
{\pi_t^\mu}_{\mu}{\pi_t^\sigma}_{\sigma}
&=&16\pi (GT_{\bot \bot}+\sigma\xi_tT_{\bot \bot}),\\
D_\mu(h_t^{-1/2}\pi_t^{\mu\nu})
&=&-8\pi(GT^\nu_{\bot}+\sigma\xi_t T^\nu_{\bot}).
\end{eqnarray}
\end{mathletters}

Consider the random process
\begin{equation}
Z_t=v(\pi_t,t).
\end{equation}
According to the Ito rule\cite{Gihman}, the SDE of the process $Z_t$ is
\begin{equation}
\label{Itoeq}
dZ_t=dv(\pi_t,t)=[\partial_tv(\pi,t)+F_t(\pi_t)\partial_{\pi}v(\pi,t)+
\frac{1}{2}B_t^2\partial_{\pi\pi}v(\pi,t)]dt+B_t\partial_{\pi}v(\pi,t)dW_t,
\end{equation}
where for simplicity we have written $\pi^{\mu\nu}, F^{\mu\nu}, B^{\mu\nu}$ and
$T^{\mu\nu}$ as $\pi$, $F$, $B$ and $T$. 

The integral form of (\ref{Itoeq}) is given by
\begin{eqnarray}
\label{integeq}
v(\pi_{t^\prime},t^\prime)-v(\pi_0,0)&=&\int^{t^\prime}_0[\partial_tv(\pi_t,t)
+F_t(\pi_t)\partial_{\pi}v(\pi_t,t)\nonumber\\
&&\mbox{}
+\frac{1}{2}B_t^2\partial_{\pi\pi}v(\pi_t,t)]dt
+\int^{t^\prime}_0B_t\partial_{\pi}v(\pi_t,t)dW_t.
\end{eqnarray}
We assume that $v(\pi,t)$ has compact support, so that
\begin{enumerate}
\item $v(\pi,0)=0$,
\item $v(\pi,\infty)=0$,
\item $\int^{\infty}_0E\{(B_t\partial_{\pi}v(\pi,t))^2\}dt < \infty$.
\end{enumerate}

We have
\begin{equation}
E\biggl\{\int^{\infty}_0B_t\partial_{\pi}v(\pi_t,t)dW_t\biggr\}=0,
\end{equation}
and for $t^\prime\rightarrow\infty$ in (\ref{integeq}) and taking the expectation value,
we get
\begin{equation}
\int^{\infty}_0E\{[\partial_tv(\pi_t,t)+F_t(\pi_t)\partial_{\pi}v(\pi_t,t)
+\frac{1}{2}B_t^2\partial_{\pi\pi}v(\pi_t,t)]\}dt=0.
\end{equation}
If the transition probability of the stochastic process $\pi_t$ possesses a density, then
we have
\begin{equation}
\int^{\infty}_0dt\int_{{\cal R}}dyp(y,t\vert \pi_t,0)[\partial_tv(y,t)+F(y)\partial_yv(y,t)
+32\pi^2 \sigma^2T^2\partial_{yy}v(y,t)]=0.
\end{equation}
By performing a partial integration, we obtain
\begin{equation}
\label{partint}
\int^{\infty}_0dt\int_{{\cal R}}dyv(y,t)[-\partial_tp(y,t\vert \pi_t,0)-
\partial_y[F_t(y)p(y,t\vert \pi_t,0)]+32\pi^2\sigma^2T^2\partial_{yy}p(y,t\vert \pi_t,0)]=0.
\end{equation}
Since $v(\pi_t,t)$ is an arbitrary function, it follows from (\ref{partint}) that
\begin{equation}
\partial_tp(y,t\vert \pi_t,0)=-\partial_y[F_t(y)p(y,t\vert \pi_t,0)]
+32\pi^2\sigma^2T^2\partial_{yy}p(y,t\vert \pi_t,0).
\end{equation}
Thus, the probability density for the random conjugate momentum variable
$\pi_t^{\mu\nu}$ satisfies a Fokker-Planck equation.

The stationary probability behavior of a gravitational system is given by
\begin{equation}
p_S(\pi)=\frac{C}{64\pi^2 T^2}
\exp\biggl(\frac{1}{32\pi^2\sigma^2}\int^\pi \frac{F(u)}{T^2}du\biggr),
\end{equation}
where $C$ is a normalization constant and we have used the Ito interpretation of
(\ref{pisde}).

\section{Stochastic Equations of Motion}

Consider the geodesic equation:
\begin{equation}
\frac{du^\mu}{ds}+\Gamma^\mu_{\alpha\beta}u^\alpha u^\beta=0,
\end{equation}
where 
\begin{equation}
\Gamma^\mu_{\alpha\beta}=\frac{1}{2}g^{\mu\sigma}
\biggl(\partial_\alpha g_{\sigma\beta}+\partial_\beta g_{\alpha\sigma}
-\partial_\sigma g_{\alpha\beta}\biggr),
\end{equation}
denotes the Christoffel symbol, $u^\mu=dx^\mu/ds$ denotes the time-like four-velocity
and $ds$ is the increment of proper time along the world line, defined by
\begin{equation}
ds^2=-g_{\mu\nu}dx^\mu dx^\nu.
\end{equation}
We can then consider $u^\mu$ to be a random variable, $u^\mu_s$, and form the SDE:
\begin{equation}
du_s^\mu+\Gamma^\mu_{s,\,\alpha\beta}u_s^\alpha
u_s^\beta ds=-\zeta_sF_s^\mu ds,
\end{equation}
where $F_s^\mu$ is a vector quantity random variable, formed from the metric tensor and its
derivatives, $\zeta_s$ is a Brownian motion process in terms of the proper time $s$, and the
Christoffel symbol is treated as a random variable determined by the stochastic
metric $g_{s,\,\mu\nu}$. At the length scale for which the 
fluctuations of spacetime are significant, we picture a test particle moving in
spacetime along a Brownian motion path such that $u_s^\mu$ does not have a
well-defined derivative with respect to $s$ at a point along the world line. For larger
macroscopic length scales for which the spacetime fluctuations can be neglected, the
motion of the test particle becomes the same as the deterministic geodesic motion in GR.

Let us define the spatial metric:
\begin{equation}
h_{\mu\nu}=g_{\mu\nu}+u_\mu u_\nu.
\end{equation}
Then, ${h^\mu}_\nu=g^{\mu\sigma}h_{\sigma\nu}$ is the projection operator onto the
subspace of the tangent space perpendicular to $u^\mu$. The expansion $\theta$, 
shear $\sigma_{\mu\nu}$ and twist $\omega_{\mu\nu}$ of a congruence of
geodesics are defined by
\begin{mathletters}
\begin{eqnarray}
\theta&=&Y^{\mu\nu}h_{\mu\nu},\\
\sigma_{\mu\nu}&=&Y_{(\mu\nu)}-\frac{1}{3}\theta h_{\mu\nu},\\
\omega_{\mu\nu}&=&Y_{[\mu\nu]},
\end{eqnarray}
\end{mathletters}
where the tensor field, $Y_{\mu\nu}$, is 
\begin{equation}
Y_{\mu\nu}=\nabla_\mu u_\nu,
\end{equation}
and $Y_{\mu\nu}$ is purely spatial:
\begin{equation}
Y_{\mu\nu}u^\mu=Y_{\mu\nu}u^\nu=0.
\end{equation}
The vector field, $u^\mu$, of tangents is normalized to unit length,
$u^\mu u_\mu=-1$, and $\nabla_\mu$ is defined by
\begin{equation}
\nabla_\mu u^\nu=\partial_\mu u^\nu+\Gamma^\nu_{\mu\sigma}u^\sigma.
\end{equation}

The Raychaudhuri equation takes the form\cite{Raychaudhuri,Hawking,Wald}
\begin{equation}
\label{Ray}
u^\sigma\nabla_\sigma\theta=\frac{d\theta}{ds}=-\frac{1}{3}\theta^2-\sigma_{\mu\nu}\sigma^{\mu\nu}+\omega_{\mu\nu}\omega^{\mu\nu}
-R_{\sigma\rho}u^\sigma u^\rho.
\end{equation}

Using Einstein's field equation, we have
\begin{equation}
R_{\mu\nu}u^\mu u^\nu=8\pi G
\biggl[T_{\mu\nu}-\frac{1}{2}g_{\mu\nu}{T^\sigma}_\sigma\biggr]u^\mu u^\nu
=8\pi G\biggl[T_{\mu\nu}u^\mu u^\nu+\frac{1}{2}{T^\sigma}_\sigma\biggr].
\end{equation}
We assume that the weak energy condition:
\begin{equation}
T_{\mu\nu}u^\mu u^\nu \geq 0
\end{equation}
holds for matter for all timelike $u^\mu$.  Moroever, we also assume that the strong
energy condition holds:
\begin{equation}
\label{strong}
T_{\mu\nu}u^\mu u^\nu \geq -\frac{1}{2}{T^\sigma}_\sigma.
\end{equation}

We choose the geodesic congruence to be hypersurface orthogonal, so that
$\omega_{\mu\nu}=0$, whereby
the third term in (\ref{Ray}) vanishes. It follows from the condition (\ref{strong})
that the left hand-side of (\ref{Ray}) is negative. Then we have
\begin{equation}
\frac{d\theta}{ds}+\frac{1}{3}\theta^2 \leq 0,
\end{equation}
which gives
\begin{equation}
\frac{d}{ds}(\theta^{-1}) \geq \frac{1}{3},
\end{equation}
and, therefore,
\begin{equation}
\label{thetabound}
\theta^{-1}(s)\geq\theta_0^{-1}+\frac{1}{3}s,
\end{equation}
where $\theta_0$ is the initial value of $\theta$. If we assume that $\theta_0$ is negative
for a converging congruence of geodesics, then (\ref{thetabound}) implies that $\theta^{-1}$
must pass through zero and $\theta\rightarrow -\infty$ within a proper time
$s\leq3/\vert\theta_0\vert$. Thus, it follows for a tube of matter that at a point where
$\theta\rightarrow-\infty$, the matter density $\rho\rightarrow\infty$ and there is
a singularity at that point on the world line. 

For null geodesics, the Raychaudhuri equation takes the form:
\begin{equation}
\label{Raynull}
\frac{d\theta}{d\lambda}=-\frac{1}{2}\theta^2
-{\hat \sigma}_{\mu\nu}{\hat\sigma}^{\mu\nu}
+{\hat\omega}_{\mu\nu}{\hat\omega}^{\mu\nu}-R_{\mu\nu}k^\mu k^\nu,
\end{equation}
where we now consider a congruence of null geodesics with the tangent field $k^\mu$.
The energy condition takes the form:
\begin{equation}
T_{\mu\nu}k^\mu k^\nu \geq 0.
\end{equation}

Let us write (\ref{Ray}) in the form:
\begin{equation}
\frac{d\theta}{ds}=-(\frac{1}{3}\theta^2+\sigma_{\mu\nu}\sigma^{\mu\nu}
+8\pi G{\tilde T}),
\end{equation}
where we have again chosen a congruence which is hypersurface orthogonal,
$\omega_{\mu\nu}=0$, and
\begin{equation}
{\tilde T}=T_{\mu\nu}u^\mu u^\nu+\frac{1}{2}{T^\sigma}_\sigma.
\end{equation}
We can now use the stochastic control parameter, $G_s=G+\sigma\xi_s$, and obtain
the SDE:
\begin{equation}
\label{Raysde}
d\theta_s=-(\frac{1}{3}\theta^2_s+\sigma_{s,\,\mu\nu}\sigma_s^{\mu\nu}
+8\pi G{\tilde T})ds-8\pi\sigma{\tilde T}dW_s,
\end{equation}
where $dW_s$ denotes the Wiener process, $dW_s=d\xi_sds$.  For
null geodesics, we obtain the SDE:
\begin{equation}
\label{Raynullsde}
d\theta_\lambda=-(\frac{1}{2}\theta_\lambda^2+{\hat\sigma}_{\lambda,\,\mu\nu}
{\hat\sigma}_\lambda^{\mu\nu}
+8\pi G{\tilde T})d\lambda -8\pi\sigma{\tilde T}dW_\lambda.
\end{equation}

We see that for a given length scale if the intensity of the fluctuations is big enough, then
the left hand-sides of (\ref{Raysde}) and (\ref{Raynullsde}) are no longer negative
definite and it no longer follows that $\vert\theta\vert\rightarrow\infty$ at points
along the world line. Caustic singularities can be prevented from developing in a congruence of
timelike or null geodesics if convergence
occurs anywhere in the manifold, provided the geometrical fluctuations
are strong enough. Thus, stochastic gravity
can avoid the occurrence of singularities in spacetime. In the limit of classical GR,
the Hawking-Penrose theorems hold, for the Brownian motion fluctuations of
spacetime are negligible and can be neglected.  

The big-bang singularity
in cosmology and the singularity which occurs in the gravitational collapse of a star,
in GR, can both be avoided in stochastic gravity. It is to be expected that in the limit
of zero spatial dimensions and zero time, the spacetime fluctuations will dominate and
smear out singularities. In the next section, we shall see that it is also possible to
avoid the occurrence of infinite red shift event horizons in the collapse of stars.

\section{Gravitational Collapse}

Let us apply our stochastic gravitational theory to the problem of gravitational collapse of 
a star. A cooling star of mass greater than the Chandrasekhar or Oppenheimer-Volkoff
mass limit cannot maintain equilibrium as either a white dwarf of a neutron star, and is
predicted in GR to collapse to a black hole. We shall restrict ourselves to the case
of spherically symmetric collapse of ``dust" with negligible pressure, In practise, we
should solve Hamilton's first order equations (\ref{heq}) and (\ref{pieq}), since
(\ref{pieq}) is linear in the control parameter $G$ and we can solve the stochastic
equation (\ref{piwiener}) directly in the Gaussian white-noise limit. However, instead
we shall solve the collapse problem by following the treatment given by
Oppenheimer and Snyder and by Weinberg\cite{Oppenheimer,Weinberg}.

We use a comoving coordinate system to
describe the freely falling dust particles. The metric is given in Gaussian normal form by
\begin{equation}
ds^2=dt^2-U(r,t)dr^2-V(r,t)(d\theta^2+\sin^2\theta d\phi^2),
\end{equation}
while the energy-momentum tensor for the fluid is
\begin{equation}
T^{\mu\nu}=\rho u^\mu u^\nu,
\end{equation}
where $\rho(r,t)$ is the proper energy density and $u^\mu$ is 
given in comoving coordinates by
\begin{equation}
u^r=u^{\theta}=u^{\phi}=0,\quad u^0=1.
\end{equation}
The energy conservation equation is
\begin{equation}
\partial_t(\rho V\sqrt{U})=0.
\end{equation}
The Einstein field equations are given by
\begin{equation}
R_{\mu\nu}=8\pi GN_{\mu\nu},
\end{equation}
where
\begin{equation}
N_{\mu\nu}=T_{\mu\nu}-\frac{1}{2}g_{\mu\nu}{T^\sigma}_\sigma
=\rho(\frac{1}{2}g_{\mu\nu}+u_\mu u_\nu).
\end{equation}

Assuming that the collapse is homogeneous, we can seek a separable solution:
\begin{equation}
U=R^2(t)f(r),\quad V=S^2(t)g(r).
\end{equation}
Then, Einstein's field equations require that\cite{Weinberg} ${\dot S}/S={\dot R}/R$,
where ${\dot R}=\partial_tR$. We can choose : $S(t)=R(t)$ and redefine the 
radial coordinate, so that $V=R^2(t)r^2$. Solving the resulting Einstein field equations
leads to the Friedmann-Robertson-Walker (FRW) metric:
\begin{equation}
\label{FRW}
ds^2=dt^2-R^2(t)\biggl[\frac{dr^2}{1-kr^2}+r^2d\theta^2
+r^2\sin^2\theta d\phi^2\biggr].
\end{equation}

We normalize $R(t)$ so that $R(0)=1$ and obtain: 
\begin{equation}
\rho(t)=\rho(0)R^{-3}(t).
\end{equation}
The field equations yield
\begin{equation}
\label{Rdoteq}
{\dot R}^2(t)=\frac{8\pi G}{3}\frac{\rho(0)}{R(t)}-k,
\end{equation}
where $k$ is a constant given by
\begin{equation}
k=\frac{8\pi G}{3}\rho(0).
\end{equation}
Eq.(\ref{Rdoteq}) becomes
\begin{equation}
\label{Rdotsquared}
{\dot R}^2=k\biggl[\frac{1}{R(t)}-1\biggr],
\end{equation}
with the parametric cycloid solution:
\begin{mathletters}
\begin{eqnarray}
t&=&\biggl(\frac{\psi+\sin\psi}{2\sqrt{k}}\biggr),\\
R&=&\frac{1}{2}(1+\cos\psi).
\end{eqnarray}
\end{mathletters}
At time $t=t_S$, where 
\begin{equation}
t_S=\frac{\pi}{2\sqrt{k}}=\frac{\pi}{2}\biggl(\frac{3}{8\pi G\rho(0)}\biggr),
\end{equation}
the fluid sphere collapses to an infinite proper energy density. The resulting
singularity as the end stage of collapse is inevitable in classical GR, as follows from the
Hawking-Penrose theorems on gravitational collapse\cite{Hawking}. 

According to the Birkhoff theorem, the metric outside the collapsing star is static
and given by the Schwarschild solution:
\begin{equation}
ds^2=\biggl(1-\frac{2GM}{r}\biggr)dt^2-\biggl(1-\frac{2GM}{r}\biggr)^{-1}dr^2
-r^2(d\theta^2+\sin^2\theta d\phi^2).
\end{equation}
By choosing an integrating factor, we can transform the metric to a standard
form\cite{Weinberg}:
\begin{equation}
ds^2=B(r,t)dt^2-A(r,t)dr^2-r^2(d\theta^2+\sin^2\theta d\phi^2),
\end{equation}
where
\begin{mathletters}
\begin{eqnarray}
B&=&\frac{R}{S}\biggl(\frac{1-kr^2}{1-ka^2}\biggr)^{1/2}
\frac{(1-ka^2/S)^2}{(1-kr^2/R)},\\
A&=&\biggl(1-\frac{kr^2}{R}\biggr)^{-1},
\end{eqnarray}
\end{mathletters}%
and the constant $a$ is equated to the radius of the sphere in comoving polar coordinates.
It follows that the interior and the exterior solutions match continuously at $r=aR(t)$
when
\begin{equation}
k=\frac{2GM}{a^3},
\end{equation}
which gives $M=\frac{4\pi}{3}\rho(0)a^3$.

Let us rewrite Eq.(\ref{Rdotsquared}) as
\begin{equation}
\label{sqrtdotR}
{\dot R}(t)=-\biggl(\frac{2GM}{a^3}\biggr)^{1/2}\biggl[\frac{1}{R(t)}-1\biggr]^{1/2},
\end{equation}
where for the collapse problem we have chosen the negative square root. The nonlinear
SDE has the form:
\begin{equation}
dR_t=[\sqrt{G}f(R_t)+\beta(\sqrt{G}+\zeta_t)f(R_t)]dt,
\end{equation}
where
\begin{equation}
f(R_t)=-\biggl(\frac{2M}{a^3}\biggr)^{1/2}\biggl(\frac{1}{R_t}-1\biggr)^{1/2}.
\end{equation}
For rapidly varying geometrical fluctuations, corresponding to a random process with a
$\zeta_t$ that has a short correlation time, we can obtain in the white-noise limit the
approximate stationary probability density:
\begin{equation}
p_S(R)=\frac{C}{v^2(R)}
\exp\biggl[\frac{2\sqrt{G}}{\sigma^2}\int^R\frac{dR}{v(R)}\biggr],
\end{equation}
where $C$ is a normalization constant and
\begin{equation}
v(R)=f(R)[1+\tau_{\rm cor}Q_1(R)+...].
\end{equation}

To lowest order we get
\begin{equation}
p_S(R)\sim C\biggl(\frac{a^3}{2M}\biggr)
\biggl(\frac{R}{1-R}\biggr)
\exp\biggl\{\frac{2\sqrt{G}}{\sigma^2}\biggl(\frac{a^3}{2M}\biggr)^{1/2}
[\sqrt{R}\sqrt{1-R}-\arcsin{\sqrt{R}}]\biggr\}.
\end{equation}
We see that $R=1$ is a natural boundary\cite{Gihman}: both the drift and the diffusion
coefficients vanish at
$R=1$. For $\sqrt{G} < \sigma^2/2(a^3/2M)^{1/2}$ the point $R=1$ is attracting and the
stationary probability
is mostly concentrated at $R=1$. The point $\sqrt{G}=\sigma^2/2(a^3/2M)^{1/2}$ is a transition
point, since 
for $\sqrt{G} > \sigma^2/2(a^3/2M)^{1/2}$ the probability density moves away from the
divergent point $R=1$
and a new stationary probability density emerges with a large finite probability for collapse
towards
$R=0$. However, for $R\rightarrow 0$, we have
\begin{equation}
p_S(R)\sim 0.
\end{equation}
We therefore arrive at the result that as the star collapses there is {\it zero
probability for $R(t)$ to have the value zero, and consequently there is zero probability
of having a singularity as the final state of collapse}.

Let us now consider the red shift emitted from the surface of the star as it collapses.
The fractional change of wave length emitted at the surface is
\begin{equation}
z=\frac{\lambda^\prime-\lambda_0}{\lambda_0}
=\frac{dt^\prime}{dt}-a{\dot R}(t)\biggl(1-\frac{2GM}{aR(t)}\biggr)^{-1}-1,
\end{equation}
where
\begin{equation}
t^\prime=t+\int^t_{aR(t)}\biggl(1-\frac{2GM}{r}\biggr)^{-1}dr
\end{equation}
is the time it takes for a light signal emitted in a radial direction at standard time $t$
to reach a distant point $r$.  In the limit, $R(t)\rightarrow 2GM/a=ka^2$, we obtain
by using (\ref{sqrtdotR}):
\begin{equation}
\label{zeq}
z=2\biggl(1-\frac{ka^2}{R(t)}\biggr)^{-1}.
\end{equation}
We have
\begin{equation}
dz=f(G,R)dt,
\end{equation}
where
\begin{eqnarray}
f(G,R)&=& -\frac{4GM}{a}\biggl(1-\frac{2GM}{aR}\biggr)^{-2}
\frac{{\dot R}}{R^2}\nonumber\\
&=&\biggl(\frac{2}{a}\biggr)^{5/2}(GM)^{3/2}\biggl(1-\frac{2GM}
{aR}\biggr)^{-2}
\frac{(1-R)^{1/2}}{R^{5/2}}.
\end{eqnarray}
We can now define the nonlinear SDE:
\begin{equation}
dz_t=\mu(G)u(R_t)dt+\gamma(G+\zeta_t)u(R_t)dt,
\end{equation}
where the function $u(R)\rightarrow\infty$ as $R\rightarrow 2GM/a$, corresponding
to the infinite red shift limit as the event horizon is approached during collapse
predicted by classical GR, and
$\zeta_t$ is a short time stochastic, colored process. Expanding around the white-noise
Gaussian limit, we can derive the approximate stationary probability density
for the red shift in the limit $R\rightarrow 2GM/a$:
\begin{equation}
\label{probability}
p_S(z)\sim\frac{C}{u^2(R)}\exp\biggl[\frac{2\mu(G)}{\sigma^2}\int\frac{dR}{u(R)}\biggr].
\end{equation}

The result emerges from (\ref{probability}) that provided the bound
\begin{equation}
\label{bound}
\int\frac{dR}{u(R)} < \frac{\sigma^2}{\mu(G)}\ln(u)
\end{equation}
is satisfied, then
\begin{equation}
p_S(z)\sim 0
\end{equation}
and the probability for an infinite red shift at $R=2GM/a$ is zero. Thus, in our stochastic
gravitational theory the probability for a black hole event horizon with an infinite red
shift to form during gravitational collapse is zero when (\ref{bound}) is satisfied. 
On the other hand, once a trapped surface forms in classical GR collapse, then the red
shift seen by a distant observer must inevitably become infinite at the radius $r_G=2GM$. 
We recall that for physical collapse in nature, the unphysical regions which occur in the Penrose
diagram for the extended (Kruskal) Schwarzschild solution are covered up by the collapsing
matter. In GR, the metric outside the collapsing star must be the Schwarzschild static metric
(Birkhoff theorem).

In stochastic gravity, the geometrical fluctuations about the deterministic event horizon cut
off the high wave length at a finite value, $\lambda=\lambda_c$, when (\ref{bound}) holds, as
viewed by an
observer at large distances from the collapsing star. In fact, the distant observer does not see a
collapsed object described by a static Schwarzschild metric with no hair. The collapsed object
will fluctuate about the average Schwarzschild solution with a characteristic time
determined by the size of the correlation function for the Schwarzschild metric.
The red shift of the collapsed object can be high, so that an outside observer would
believe that it is a black hole. The cooperative effects associated with the
self-organising microscopic subsystems comprising the event horizon can produce a large or
even infinite correlation length for the microscopic fluctuations, cutting off the infinite
wavelength radiation emitted by the macroscopic surface of the collapsing star as $R\rightarrow
2GM/a$.

Our use of the stationary probability density for the collapse problem means that we
have assumed that after an infinite time has elapsed the system evolves to a
stationary state. This is compatible with the fact that the spherically symmetric
collapse problem has well-defined flat space asymptotic limits as $r\rightarrow\infty$.
Since in stochastic gravity, it can take a long (but finite) time for a distant observer to see the
formation of an apparent event horizon during collapse, this is consistent with the observer
measuring the stationary probability density $p_S$ as the limit $p(z)\rightarrow p_S(z)$ as
$t\rightarrow\infty$.

We have succeeded in removing both the singularity in collapse and the black
hole event horizon in our stochastic probability gravitational theory (provided 
(\ref{bound}) is satisfied.) The 
deterministic GR predictions are clearly obtained in the limit that the intensity of
fluctuations vanishes. As we have seen in the last section, the results are consistent
with the Hawking-Penrose
theorems for collapse, because for large enough fluctuations of the stochastic metric,
$g_{s,\,\mu\nu}$, caustics can be prevented from occurring
for converging congruences of timelike or null geodesics in the spacetime manifold.

The `greyness' of the dense collapsed object depends on the size of the correlation time
$\tau_{\rm cor}$. The correlation length can be significantly larger than the Planck length
scale, $l_P$, since
the fluctuations are associated with a stochastic property of the gravitational field,
which is not directly proportional to Planck's constant $\hbar$, and is therefore
not strongly correlated to regions at the scale of $l_P$.
If the dense collapsed object is sufficiently grey, then this could solve the
information loss problem first introduced by Hawking\cite{Hawking2}. Again, this
depends on the size of the correlation length associated with the geometrical fluctuations.
Since the deterministic event horizon has been removed, either there is no Hawking radiation
emitted from the surface of the collapsed object, or the spectrum of the radiation is significantly
modified from its behavor in deterministic GR. This raises the question: will black holes
form in the collapse of astrophysical objects in our statistical mechanical treatment of 
gravity? If the answer is no, then what kind of object would describe the final state of 
collapse? These interesting questions require further investigation, before
any conclusions can be reached about this fundamental problem.

\section{The Cosmological Big-Bang Singularity}

Let us now apply our stochastic formalism to cosmology. We shall consider the
standard FRW cosmology. The FRW metric has the
form (\ref{FRW}) and solving Einstein's field equations including the cosmological
constant term yields the Friedmann equation:
\begin{equation}
\label{Friedmann}
{\dot R}^2(t)=\frac{8\pi G}{3}R^2(t)\biggl(\frac{\rho_{0,M}}{R^3(t)}
+\frac{\rho_{0,r}}{R^4(t)}\biggr)+\frac{\Lambda}{3}R^2(t)-k,
\end{equation}
where $R(t)$ denotes the cosmic scale factor, $k$ now denotes the measure of
the spatial curvature: $k=1,0,-1$, $\rho_{0,M}$ and $\rho_{0,r}$ denote the
present density of matter and radiation, respectively, and
$\Lambda$ denotes the cosmological constant. 

Let us consider for simplicity the case of zero spatial curvature and zero
cosmological constant, $\Lambda=k=0$. We shall focus on the radiation dominated
universe near $t=0$. Then, Eq.(\ref{Friedmann}) can be written as
\begin{equation}
dR=\sqrt{G}\biggl(\frac{8\pi\rho_{0,r}}{3}\biggr)^{1/2}\frac{1}{R^2}dt.
\end{equation}
We can now form the nonlinear SDE:
\begin{equation}
dR_t=\sqrt{G}f(R_t)dt+\alpha(\sqrt{G}+\zeta_t)f(R_t)dt,
\end{equation}
where $\zeta_t$ is assumed to be a short correlation time stochastic process associated
with colored fluctuations of the geometry, and
\begin{equation}
f(R_t)=\biggl(\frac{8\pi\rho_{0,r}}{3}\biggr)^{1/2}\frac{1}{R_t^2}.
\end{equation}

The approximate stationary probability density calculated in the limit of white-noise is
given by
\begin{equation}
p_S(R)\sim C\biggl(\frac{3R^4}{8\pi\rho_{0,r}}\biggr)
\exp\biggl[\biggl(\frac{2\sqrt{G}}{\sigma^2}\biggr)\biggl(\frac{1}{24\pi\rho_{0,r}}
\biggr)^{1/2}R^3\biggr],
\end{equation}
where $C$ is a normalization constant.
We have $R\rightarrow 0$ as $t\rightarrow 0$ and we see that the probability of
$R$ reaching $t=0$ is zero.  Thus, in our stochastic gravity the probability of the
big-bang singularity occurring is zero. As in the case of gravitational collapse, the
fluctuations of the geometry near $t=0$ smear out the singularity. The result is
consistent with the Hawking-Penrose theorem\cite{Hawking}, which states that in
classical GR, the universe must collapse to a singularity at $t=0$ provided the positive
energy conditions are satisfied, because for large enough fluctuations in the
neighborhood of $t=0$, the converging geodesic congruences can be prevented from
forming a singularity, as has been demonstrated in section VII.

\section{Conclusions}

We have formulated a self-consistent gravitational theory based on stochastic diffusion
processes. The geometry of spacetime is assumed to be like a fluctuating
environment with the control parameter, $G_x=G+\sigma\xi_x$. The fluctuations of the
control parameter
form a bell-shaped distribution about the deterministic Newtonian gravitational
constant $G$. This theory represents a statistical mechanical treatment of gravity
with a clearly determined classical GR limit when the intensity of the spacetime
fluctuations vanishes. In this approach, Einstein's gravitational theory is a
deterministic ``effective" theory, which holds to a very good approximation for
macroscopic gravitational systems with scales larger than the correlation lengths
associated with the geometrical fluctuations. The stochastic gravity theory does not
represent a complete
quantum mechanical theory and cannot solve the problem of quantum gravity,
but it does lead to a methodology that can account for the complexity of gravitational
phenomena as are experienced in cosmology, gravitational collapse
and astrophysics. Since it is a probabilistic theory, it reconciles a probabilistic
interpretation of the energy-stress tensor of matter and the geometry of spacetime.
It leads to critical phenomena and self-organization of gravitational systems, which
are strongly expected to play a significant role in astrophysics and cosmology\cite{Moffat5}.

By working with a $(3+1)$ foliation of spacetime, and using the
Arnowitt, Deser and  Misner formalism, we were able to obtain a stochastic differential
equation for the random canonically conjugate momentum $\pi_t^{\mu\nu}$.
We also derived a Fokker-Planck equation for the probability density of
$\pi_t^{\mu\nu}$ and showed how a stationary probability density for the
dynamical variable $\pi_t^{\mu\nu}$ could be obtained.

Stochastic geodesic congruences were studied and from the Raychaudhuri equation,
we derived a stochastic differential equation for the expansion parameter $\theta$,
which led to the result that, assuming the weak and strong positive energy conditions
are satisfied and for sufficiently strong spacetime fluctuations
at small scales, a congruence of converging geodesics can be prevented from forming caustic
singularities in the manifold, thereby avoiding the classical GR Hawking-Penrose
singularity theorems.

For certain applications of the formalism, we found it necessary to consider nonlinear
non-Markovian colored processes which were expanded about the white-noise Gaussian
limit, a limit expected to hold for a rapidly fluctuating spacetime geometry.
Stationary state probability densities were obtained in
the Gaussian limit with a short correlation time for the fluctuating geometry
in gravitational collapse and cosmology. It was found that the
fluctuations of the geometry smeared out the singularity in gravitational collapse.
Similarly, the big-bang singularity was smeared out in a probabilistic sense, leading to a
singularity-free FRW universe.

A stochastic treatment of gravitational collapse
resulted in the removal of black hole event horizons (given that the bound
(\ref{bound}) is satisfied) by cutting off the high frequencies in
the infinite red shift limit. This means that
either there is no Hawking radiation emitted from the surfaces of the collapsed objects,
or the blackbody radiation spectrum predicted by deterministic GR is
significantly modified.
Moreover, the lack of a rigorous event horizon could remove the problem of information 
loss\cite{Hawking2} depending on the `greyness' of the collapsed objects. 

We can also extend the stochastic methods developed here to the nonsymmetric
gravitational theory (NGT)\cite{Moffat6}, so that for small enough fluctuations of the
non-Riemannian geometry we obtain the classical NGT theory.

\acknowledgments
I thank M. Clayton, N. Cornish, L. Demopolous, J. L\'egar\'e, 
J. Preskill, and P. Savaria for helpful and stimulating discussions. This work
was supported by the Natural Sciences and Engineering Research
Council of Canada.

\end{document}